\documentclass[12pt]{article}
\def\phi{\varphi}

\def\d{\delta}

\def\l{\lambda}

\def\dd{\partial}

\def\one#1{#1^{\raise5pt\hbox{$\scriptstyle\!\!\!\!1$}}\,{}}
\def\two#1{#1^{\raise5pt\hbox{$\scriptstyle\!\!\!\!2$}}\,{}}
\def\three#1{#1^{\raise5pt\hbox{$\scriptstyle\!\!\!\!3$}}\,{}}

\def\id{\hbox{{1}\kern-.25em\hbox{\rm l}}}

\def\beq{\begin{equation}}
\def\eeq{\end{equation}}
\def\be{\begin{displaymath}}
\def\ee{\end{displaymath}}
\def\bea{\begin{eqnarray}}
\def\eea{\end{eqnarray}}
\def\beas{\begin{eqnarray*}}
\def\eeas{\end{eqnarray*}}
\def\bds{\begin{description}}
\def\eds{\end{description}}
\def\bmat{\left(\begin{array}}
\def\emat{\end{array}\right)}

\def\Ref#1{(\ref{#1})}
\def\?{(?)\marginpar{|?}}

\newfont{\bbd}{msbm10 scaled\magstep1} 
\def\C{\hbox{\bbd C}}                  
\def\P{\hbox{\bbd P}}                  

\renewcommand{\theequation}{\thesection.\arabic{equation}}
\newcounter{subequation}[equation]
\makeatletter
 
\expandafter\let\expandafter
\reset@font\csname reset@font\endcsname
 
\def\subeqnarray{\arraycolsep1pt
    \def\@eqnnum\stepcounter##1{\stepcounter{subequation}%
        {\reset@font\rm(\theequation\alph{subequation})}}
\jot5mm     \eqnarray}

\makeatother
\newcommand{\newsection}[1]{
\vspace{10mm}
\pagebreak[3]
\refstepcounter{section}
\setcounter{equation}{0}
\setcounter{subsection}{0}
\setcounter{footnote}{0}
 
\begin{flushleft}
{\Large\bf \thesection. #1}
\end{flushleft}
\nopagebreak
\medskip
\nopagebreak}

\newcommand\ftnote[1]{\setcounter{footnote}{#1}\addtocounter{footnote}{-1}
\footnote}
\def\JPA{J.\ Phys.\ A: Math.\ Gen.\ }
\def\one#1{#1^{\raise5pt\hbox{$\scriptstyle\!\!\!\!1$}}}
\def\two#1{#1^{\raise5pt\hbox{$\scriptstyle\!\!\!\!2$}}}
\def\P{{\cal P}}
\def\id{\hbox{1\hskip-3pt{\rm l}}}
\def\bt{B\"acklund transformation}

\def\RR{{\cal R}}

\begin{document}
\begin{flushright}
\sf LPENSL-Th 06/99 \\
\sf solv-int/9903017 \\
\end{flushright}
\vskip1cm
\begin{center}\LARGE\bf
Canonicity of \bt:
$r$-matrix approach. II.
\end{center}
\vskip1cm
\begin{center}
E K Sklyanin\ftnote{1}{On leave from: Steklov Mathematical Institute at
St.~Petersburg, Fontanka 27, St.~Petersburg 191011, Russia. 
E-mail: {\tt sklyanin\symbol{'100}euclid.pdmi.ras.ru}}
\vskip0.4cm
Laboratoire de Physique\ftnote{2}{UMR 5672 du CNRS et de l'ENS Lyon}, 
Groupe de Physique Th\'eorique, ENS Lyon, \\
46 all\'ee d'Italie, 69364 Lyon 07, France
\end{center}
\vskip2cm
{\bf Abstract.} This is the second part of the paper devoted to the general
proof of canonicity of \bt\ (BT) for a Hamiltonian
integrable system governed by $SL(2)$-invariant $r$-matrix. Introducing an
extended phase space from which the original one is obtained by imposing a
1st kind constraint, we are able to prove the canonicity of BT in a new way.
The new proof allows to explain naturally the  fact why the gauge 
transformation matrix $M$ associated to the BT has the same structure as 
the Lax operator $L$. The technique is illustrated on the example of the DST 
chain.
\vskip1cm
\begin{center}
 25 March, 1999
\end{center}

\vskip2cm
To be published in: {\sl Trudy MIAN, v. 226 (1999), Moskow, Nauka}

\newpage

\newsection{Introduction}

In the first part of this paper \cite{Skl51} we have addressed the problem
of proving the canonicity of \bt\ for the Hamiltonian integrable models
associated to $SL(2)$-invariant $r$-matrices (such as Heisenberg magnet,
Toda lattice, Nonlinear Schr\"odinger equation). 

Consider a finite-dimensional Hamiltonian system 
defined in terms of canonical variables $(X,x)$:
\beq
  \{X_i,X_j\}=\{x_i,x_j\}=0, \qquad
  \{X_i,x_j\}=\d_{ij}.
\eeq

Suppose that the system is completely integrable and possesses
a Lax matrix $L(u;X,x)$ 
whose spectral invariants generate the commuting
Hamiltonians $H_n$ ($u$ being a complex parameter).
In what follows we often shall make use of several sets of canonical
variables, like $(X,x)$, $(Y,y)$, $(S,s)$, $(T,t)$ etc.
To simplify the notation, we shall mark the Lax matrix expressed in
terms of the corresponding variables as, for example, 
$L^{(x)}(u)\equiv L(u;X,x)$ omitting the subscript when it is not important.

As in \cite{Skl51}, we assume that $L(u)$ is a matrix of order
$2\times2$ and the Poisson brackets between the entries of $L(u)$ can be
expressed in the $r$-matrix form \cite{FT87}
\beq
 \{\one L(u),\two L(v)\}=[r_{12}(u-v),\one L(u)\two L(v)]
\label{pb-LL}
\eeq
where
$\one L=L\otimes\id$, $\two L=\id\otimes L$, and
$ r_{12}=\kappa(u-v)^{-1}\P_{12}$
is the standard $SL(2)$-invariant solution to the classical Yang-Baxter 
equation \cite{FT87}, 
$\kappa$ being a constant and $\P_{12}$ being the
permutation operator in $\C^2\otimes\C^2$.

A \bt, by definition \cite{KS5}, is a one-parametric family ${\cal B}_\l$
of commuting ${\cal B}_\l\circ{\cal B}_\mu={\cal B}_\mu\circ{\cal B}_\l$
$\forall\l,\mu\in\C$
canonical transformations ${\cal B}_\l:(X,x)\rightarrow(Y,y)$
preserving the Hamiltonians $H_n$. Since the Hamiltonians, in our case,
are the spectral invariants of $L(u)$ there must exist a matrix
$M_\l(u)$ such that
\beq
 M_\l(u)L^{(x)}(u)=L^{(y)}(u)M_\l(u).
\eeq

Here it is assumed  that $Y=Y(X,x)$ and $y=y(X,x)$, so $M_\l(u)=M_\l(u;X,x)$.
Typically, $M_\l(u)$ depends on the difference $u-\l$ which we shall assume
henceforth.

In \cite{Skl51} and in the present paper we study, so to say, inverse problem,
namely, given an integrable system and a Lax matrix, find such a matrix
$M(u)$ that the similarity transformation
\beq
 M_(u-\l)L^{(x)}(u)=L^{(y)}(u)M_(u-\l)
\label{eq:ML=LM}
\eeq
induces a canonical transformation ${\cal B}_\l:(X,x)\rightarrow (Y,y)$.

In \cite{Skl51} we have proved by a direct calculation the canonicity
of the two-parametric transformation ${\cal B}_{\l_1\l_2}$ 
corresponding to the matrix $M_{\l_1\l_2}(u)$ of the form
\beq
   M_{\l_1\l_2}(u)=\left(\begin{array}{cc}
         u-\l_1+pq & p \\ -pq^2+2\mu q & u-\l_2-pq
         \end{array}\right).
\label{def-Mxxx}
\eeq

The proof, though straightforward and simple, has left one question
unanswered, namely: why the ansatz \Ref{def-Mxxx} for $M_{\l_1\l_2}(u)$
coincides with the elementary Lax matrix $L(u)$ for Heisenberg $XXX$
magnet \cite{FT87} having the Poisson brackets \Ref{pb-LL} provided
the variables $p$ and $q$ are canonical (note that in $M_{\l_1\l_2}(u)$
the variables $pq$, as shown in \cite{Skl51}, {\it are not} canonical
but have rather complicated brackets.

In the present paper we resolve this mistery and show that the abovementioned
similarity between $M(u)$ and $L(u)$ is by no means accidental but,
on the contrary, quite natural. The idea is to introduce an
extended phase space from which the original one is obtained by imposing a
1st kind constraint. Such an indirect approach turns out not only to be
simpler than the straightforward method of \cite{Skl51} but allows also 
to construct in a systematic way multiparametric families of
\bt s.

In the next section we give the general construction and a proof
of canonicity of \bt s. In section 3 we modify our construction to 
the case of integrable chains and in section 4 illustrate it on the
example of the so-called DST chain (a degeneration of isotropic Heisenberg
magnet).

\newsection{General scheme}

The first step of our construction is to extend the phase space
by supplementing the original canonical variables $(X,x)$ with
an independent set of auxiliary canonical variables $(S,s)$. Note
that the variables $(S,s)$ commute with $(X,x)$ and the dimension of
the space $(S,s)$ has no relation to the dimension of $(X,x)$.
 
Let matrix $M^{(s)}(u)$ has the same Poisson brackets \Ref{pb-LL}
as $L(u)$
\beq
  \{\one M(u),\two M(v)\}=[r_{12}(u-v),\one M(u)\two M(v)].
\label{eq:pb-MM}
\eeq

Suppose that there exists a  canonical transformation
$\RR_\l:(X,x;S,s)\rightarrow(Y,y,T,t)$, determined from the equation
\beq
  M^{(s)}(u-\l)L^{(x)}(u)=L^{(y)}(u)M^{(t)}(u-\l)
\label{eq:sx=yt}
\eeq
and having the generating function $F_\l(y,t;x,s)$
\beq
 X=\frac{\dd F_\l}{\dd x}, \quad
 Y=-\frac{\dd F_\l}{\dd y}, \quad
 S=\frac{\dd F_\l}{\dd s},\quad
 T=-\frac{\dd F_\l}{\dd t}
\label{eq:XYST}
\eeq
(for simplicity, we omit the indices $i$ in $X_i$, $x_i$ etc.)

Let us impose now the constraint
\beq
  t=s, \qquad T=S.
\label{eq:constraint}
\eeq

Suppose that one can resolve the equations \Ref{eq:XYST} and 
\Ref{eq:constraint}
with respect to $s=t$ and express $X$ and $Y$ as functions of $(x,y)$. 

{\bf Proposition}. The resulting transformation $B_\l(X,x)\rightarrow(Y,y)$  
is canonical and is given by the generating function
$\Phi_\l(x,y)=F_\l\bigl(y,s(x,y);x,s(x,y)\bigr)$, such that
\beq
  X=\frac{\dd \Phi_\l}{\dd x}, \quad
 Y=-\frac{\dd \Phi_\l}{\dd y}.
\eeq

{\bf Proof.} Let  $|_{st}$ mean the restriction on the constraint
manifold $s=t=s(x,y)$. The proof consists of two lines:
\bea
  X&=&\frac{\dd\Phi_\l}{\dd x}=
  \left.\frac{\dd F_\l}{\dd x}\right|_{st}
  +\left.\frac{\dd F_\l}{\dd s}\right|_{st}\frac{\dd s}{\dd x}
  +\left.\frac{\dd F_\l}{\dd t}\right|_{st}\frac{\dd t}{\dd x} \nonumber\\
  &=&\left.\frac{\dd F_\l}{\dd x}\right|_{st}
  +\frac{\dd s}{\dd x}
  \left.\left(\frac{\dd F_\l}{\dd s}+\frac{\dd F_\l}{\dd t}\right)\right|_{st}.
\eea

We observe now that
\beq
  \left.\left(\frac{\dd F_\l}{\dd s}+\frac{\dd F_\l}{\dd t}\right)\right|_{st}
=0
\eeq
due to $S=T$, and, consequently, $X=\dd\Phi_\l/\dd x$. Similarly, one
proves  $Y=-\dd\Phi_\l/\dd y$. 

To conclude, we notice that the constraint \Ref{eq:constraint}
implies the identity $M^{(s)}(u-\l)=M^{(t)}(u-\l)$ and, consequently,
the equality \Ref{eq:sx=yt} turns into the equality \Ref{eq:ML=LM}
ensuring thus that the transformation ${\cal B}_\l$ preserves the 
spectrum of $L(u)$.

\newsection{Modification for the chain}

In many applications the Lax matrix $L(u)$ is a monodromy matrix
\cite{FT87} factorized into the product of local Lax matrices $\ell_i(u)$
\beq
 L(u)=\ell_N(u)\ldots \ell_2(u)\ell_1(u)
\eeq
having the same Poisson brackets \Ref{pb-LL} as $L(u)$
\beq
  \{\one \ell_i(u),\two \ell_j(v)\}=
[r_{12}(u-v),\one \ell_i(u)\two \ell_j(v)]\d_{ij}.
\label{eq:pb-ll}
\eeq

The similarity transformation \Ref{eq:ML=LM} is replaced now with
a gauge transformation:
\beq
   M_i(u-\l)\ell_i^{(x)}(u)=\ell_i^{(y)}(u)M_{i-1}(u-\l)
\label{eq:gauge}
\eeq
which ensures the preservation of the spectral invariants of $L(u)$
due to $M_N(u-\l)L^{(x)}(u)=L^{(y)}(u)M_N(u-\l)$.

The modification of the reduction procedure described in the previous
section is quite straightforward. Supposing that $\ell_i(u)$ and
$M_i(u)$ depend on local canonical variables
\begin{subeqnarray}
&& \ell_i^{(x)}(u)\equiv \ell_i(u;X_i,x_i), \qquad
 \ell_i^{(y)}(u)\equiv \ell_i(u;Y_i,y_i),  \\
&& M_i^{(s)}(u)\equiv M(u;S_i,s_i), \qquad
   M_i^{(t)}(u)\equiv M(u;T_i,t_i),
\end{subeqnarray}
we define first the local canonical transformations 
$\RR^{(i)}_\l:(X_i,x_i;S_i,s_i)\rightarrow(Y_i,y_i,\allowbreak T_i,t_i)$
from the equations
\beq
  M_i^{(s)}(u-\l)\ell_i^{(x)}(u)=\ell_i^{(y)}(u)M_i^{(t)}(u-\l).
\label{eq:Ml=lM}
\eeq

Let the corresponding generating functions are $f^{(i)}_\l(y_i,t_i;x_i,s_i)$.
Consider the direct product of  $N$ phase spaces
$(X_i,x_i;S_i,s_i)$ and, respectively, $(Y_i,y_i;\allowbreak T_i,t_i)$.
The generating function
\beq
  F_\l:=\sum_{i=1}^N f^{(i)}_\l(y_i,t_i;x_i,s_i)
\eeq
determines then the direct product  $\RR_\l$ of the local
canonical transformations $\RR^{(i)}_\l$.

Let us now impose the constraint
\beq
  t_i=s_{i-1}, \qquad T_i=S_{i-1}
\label{eq:constr-chain}
\eeq
assuming periodicity  $i+N\equiv i$. The proof of the canonicity of the
resulting transformation ${\cal B}_\l:(X,x)\rightarrow(Y,y)$
parallels the proof given in the previous section. It remains to
notice that after imposing the constraint \Ref{eq:constr-chain}
we have $M_i^{(s)}(u-\l)=M_i^{(t)}(u-\l)$ and obtain the equality 
\Ref{eq:gauge}.

\newsection{Classical $DST$-model}

We shall illustrate the general scheme described above
on the simple example of the so-called $DST$-model ---
a degeneration of the Heisenberg spin chain \cite{FT87}. In the notation
of the previous section, the model is characterized by the local Lax matrices
\beq
  \ell_i^{(x)}(u)=\left(\begin{array}{cc}u-c_i-X_ix_i & x_i \\ -X_i & 1
               \end{array}\right)
\eeq
whose Poisson brackets \Ref{eq:pb-ll} are governed by the standard 
$SL(2)$-invariant $r$-matrix \cite{FT87}
\beq
  r_{12}(u)=-{\cal P}_{12}u^{-1},
\label{eq:r}
\eeq
where ${\cal P}_{12}$ is the permutation operator in $\C^2\otimes\C^2$.

Let us choose for $M_i^{(s)}(u)$ another realization of the same
Poisson brackets
\beq
  M_i^{(s)}(u)=\left(\begin{array}{cc}1 & -S_i \\ s_i & u-S_i s_i
               \end{array}\right).
\label{eq:def-Mdst}
\eeq

Note that our choice \Ref{eq:def-Mdst} of ansatz for  
$M_i^{(s)}(u)$ is different from \Ref{def-Mxxx}
used in our previous paper \cite{Skl51}. In a sense, \Ref{eq:def-Mdst}
is the simplest possible choice since $\det M(u-\l)=u-\l$ has only one
pole instead of two as for \Ref{def-Mxxx}.

From \Ref{eq:Ml=lM} one easily finds the local canonical transformation 
$\RR_\l$ which is given by the equations
\begin{subeqnarray}
 X&=&s,\qquad  Y=t+\frac{(c-\l)s}{1-sy}, \\
 T&=&y, \qquad
S=x-\frac{(c-\l)y}{1-sy},
\label{eq:DST-XYST}
\end{subeqnarray}
or, equivalently, by the generating function
\beq
  f_\l(yt\mid xs) = xs-yt+(c-\l)\ln(1-sy).
\label{eq:DST-f}
\eeq

Following the general scheme, we impose now the constraint 
\Ref{eq:constr-chain}. Using (\ref{eq:DST-XYST}b) and resolving the equations 
$S_i=T_i$ with respect to $s_i$ one obtains
\beq
 s_i=\frac{1}{y_i}+\frac{c_i-\l}{y_{i+1}-x_i}
\eeq
and, after substitution to (\ref{eq:DST-XYST}a),
\begin{subeqnarray}
 X_i&=&\frac{1}{y_i}+\frac{c_i-\l}{y_{i+1}-x_i}, \\
 Y_i&=&\frac{1}{y_{i-1}}+\frac{x_i-y_{i+1}}{y_i^2}
  +\frac{c_{i-1}-\l}{y_i-x_{i-1}}-\frac{c_i-\l}{y_i}.
\label{eq:DST-Bckl}
\end{subeqnarray}

It is easy to verify that the equations \Ref{eq:DST-Bckl} describe
a canonical transformation whose generating function is
\beq
 \Phi_\l(y\mid x)=\sum_i \frac{x_i-y_{i+1}}{y_i}
     +(c_i-\l)\ln\frac{y_i}{y_{i+1}-x_i}.
\eeq

As expected from the general theory, the generating function $\Phi_\l(y\mid x)$
can also be obtained, up to a  constant term, from $F_\l$ corresponding to
\Ref{eq:DST-f} by the substitution of $s_i$ and $t_i$. 

From the general proof given in sections 2 and 3 it follows that
taking as ansatz for $M_i(u)$ any realization of the Poisson brackets
\Ref{eq:pb-MM} we will obtain some \bt. In particular, we can take for
$M_i(u)$ a product of elementary matrices like \Ref{eq:def-Mdst}.
The corresponding \bt\ will be obviously the composition of the
elementary \bt s corresponding to \Ref{eq:def-Mdst}. 
For instance, in such a way
one can recover the matrix \Ref{def-Mxxx} and the corresponding
two-parametric \bt\ ${\cal B}_{\l_1\l_2}$ from \cite{Skl51}.

\section{Discussion}

Our exposition was  very sketchy, and we are planning to publish a more
detailed study of our new technique elsewhere. Here we remark only
that applicability of our construction goes far beyond the integrable
systems connected to the $SL(2)$-invariant $r$-matrice \Ref{eq:r}.
The proof given in section 2 works for any solution to the classical
Yang-Baxter equation provided the existence of the local canonical
transformation $\RR_\l$ and the conditions of degeneracy. The work on
the application of the reduction technique to other integrable systems
is in progress.

Another important remark concerns the connection with the quantum case.
As noticed first in \cite{PG92}, see also \cite{KS5}, the quantum analog 
of \bt\ is the famous Baxter's $Q$-operator \cite{Bax82}.
It can be shown that our construction of ${\cal B}_\l$ is the exact
classical analog of the quantum construction of $Q(u)$ as a trace of
a quantum monodromy matrix. Using this analogy one can, for example,
construct the $Q$-operator for the quantum $DST$ model \cite{KSS98}.

\section*{Acknowledgements}

I am grateful to V.~B.~Kuznetsov and F.~Nijhoff for
their interest in the work and useful discussions. This work was started
during my stay at the Department of Mathematics, the University of Leeds
and benefited from the financial support of EPSRC.


\end{document}